\documentclass[fleqn,12pt]{article}
\usepackage{times}
\usepackage{geometry}
\usepackage{epsfig}
\usepackage{setspace}
\doublespace
\begin{document}                
\title{ Wide angle near-field diffraction and Wigner distribution }
\author{Jos\'e B. Almeida\\
\small{\emph{Universidade do Minho, Physics Department, 4710-057
Braga, Portugal}}\\ Vasudevan Lakshminarayanan\\
\small{\emph{University of Missouri - St. Louis, School of
Optometry}} \\ [-5mm] \small{\emph{and Department of Physics and
Astronomy, St. Louis, MO 63121, USA}}}
\date{}

%
\maketitle
\begin{abstract}                
Free-space propagation can be described as a shearing of the
Wigner distribution function in the spatial coordinate; this
shearing is linear in paraxial approximation but assumes a more
complex shape for wide-angle propagation. Integration in the
frequency domain allows the determination of near-field
diffraction, leading to the well known Fresnel diffraction when
small angles are considered and allowing exact prediction of
wide-angle diffraction. The authors use this technique to
demonstrate evanescent wave formation and diffraction elimination
for very small apertures.
\end{abstract}

The Wigner distribution function (WDF) provides a convenient way
to describe an optical signal in space and spatial frequency
\cite{Bastiaans78, Dragoman97}. The propagation of an optical
signal through first-order optical systems is well described by
the WDF transformations \cite{Bastiaans79, Bastiaans91}, which
allows the reconstruction of the propagated signal. On the other
hand, some authors have linked Fresnel diffraction and the
fractional Fourier transform (FRFT) \cite{Alieva94, Agarwal94};
both of these papers associate free-space propagation to a
rotation of the WDF and the corresponding FRFT accompanied by a
quadratic phase factor.

In this paper we show that free-space propagation is always
associated with shearing of the WDF. This can be used to evaluate
the near-field diffraction. In the paraxial approximation our
results duplicate the well known Fresnel diffraction, with the
advantage that we do not need to resort to the Cornu integrals.
The same procedure can be extended to wide angles, where other
phenomena are apparent, namely the presence of evanescent waves.

The Wigner distribution function (WDF) of a scalar, time
harmonic, and coherent field distribution $\varphi(\mathbf{q},z)$
can be defined at a $z=\mathrm{const.}$ plane in terms of either
the field distribution or its Fourier transform
$\overline{\varphi}(\mathbf{p})=\int
\varphi(\mathbf{q})\exp(-ik\mathbf{q}^T \mathbf{p})d\mathbf{q}$
\cite{Dragoman97, Bastiaans79}:
\begin{eqnarray}
 \label{eq:Wigner}
 W(\mathbf{q},\mathbf{p})&=&\int
 \varphi\left(\mathbf{q}+\frac{\mathbf{q}'}{2}\right)
 \varphi^*\left(\mathbf{q}-\frac{\mathbf{q}'}{2}\right)
 \exp\left(-i k \mathbf{q}'^T \mathbf{p}\right)d \mathbf{q}' \\
 &=& \frac{k^2}{4\pi^2}\int
 \overline{\varphi}\left(\mathbf{p}+\frac{\mathbf{p}'}{2}\right)
 \overline{\varphi}^*\left(\mathbf{p}-\frac{\mathbf{p}'}{2}\right)
 \exp\left(i k \mathbf{q}^T \mathbf{p}'\right)d \mathbf{p}'~,
\end{eqnarray}
where $\mathbf{q}$ is the position vector, $\mathbf{p}$ the
conjugate momentum, $k=2 \pi/ \lambda$ and $^*$ indicates complex
conjugate.

In the paraxial approximation, propagation in a homogeneous
medium of refractive index $n$ produces a perfect mapping of the
WDF according to the relation \cite{Dragoman97, Bastiaans79}
\begin{equation}
    \label{eq:paraxprop}
    W(\mathbf{q},\mathbf{p},z)=W(\mathbf{q}- \frac{z}{n} \mathbf{p},
    \mathbf{p},0).
\end{equation}
After the WDF has been propagated over a distance, the field
distribution can be recovered by \cite{Dragoman97}
\begin{equation}
 \label{eq:field}
 \varphi(\mathbf{q},z)\varphi^*(0,z)=\frac{1}{4 \pi^2}\int
 W(\mathbf{q}/2,\mathbf{p},z) \exp(i \mathbf{q}\mathbf{p})d
 \mathbf{p}.
\end{equation}
The field intensity distribution can also be found by
\begin{equation}
 \label{eq:intens}
 |\varphi(\mathbf{q},z)|^2= \frac{4 \pi^2}{k^2}\int W(\mathbf{q},\mathbf{p},z) d
 \mathbf{p}.
\end{equation}

\begin{figure}[p]
\caption{Fresnel diffraction pattern for a one-dimensional
aperture of width $0.1~\mathrm{mm}$ with $k=10^7$.}
    \vspace{20mm}
    \centerline{\psfig{file=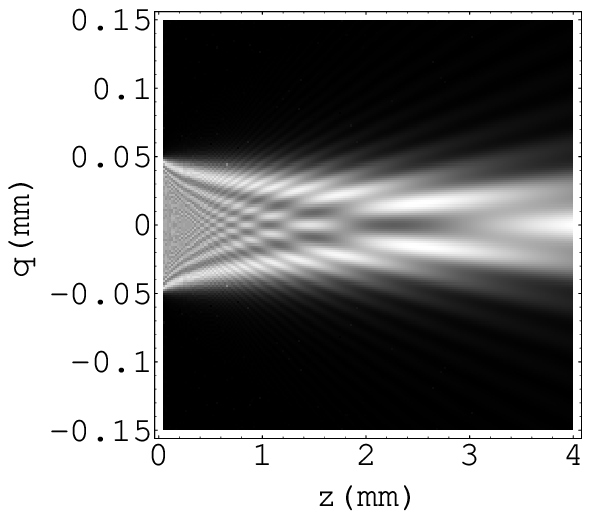, scale=2}}
\label{fig:fresnel}
\end{figure}

Eqs. (\ref{eq:field}) and (\ref{eq:intens}) are all that is
needed for the evaluation of Fresnel diffraction fields. Consider
the diffraction pattern for a rectangular aperture in one
dimension. The WDF of the aperture is given by
\begin{equation}
   \label{eq:aperture}
   W(q,p)= \left\{\begin{array}{lc}
     0 & ~~ |q|\geq l/2 \\
     2\sin [k p (l - 2q)]/k p & ~~ 0 \leq q < l/2 \\
     2\sin [k p (l + 2q)]/k p & ~~ -l/2 \leq q \leq 0 \
   \end{array} \right. ,
\end{equation}
with $l$ being the aperture width. After propagation and
integration in $p$ we obtain
\begin{eqnarray}
 \label{eq:fresnel}
 \frac{k^2|\varphi(q,z)|^2}{8 \pi^2} &=& \int_{n( 2q-l)/(2z)}^{n q/z}
 \frac{1}{k p} \sin [k p (l - 2 z p/n - 2 q)]d p  \nonumber\\
 &&  + \int^{n(2q+l)/(2z)}_{n q/z}  \frac{1}{k p} \sin [k p
 (l + 2 z p/n + 2 q)]d p .
\end{eqnarray}
Fig.\ \ref{fig:fresnel} shows a typical diffraction pattern
obtained in this way.

For wide angles the WDF can no longer be mapped from the initial
to the final planes. It must be born in mind that the components
of the conjugate momentum $\mathbf{p}$ correspond to the
direction cosines of the propagating direction multiplied by the
refractive index; the appropriate transformation is given by
\cite{Wolf99}
\begin{equation}
    \label{eq:wideshear}
    W(\mathbf{q},\mathbf{p},z)= \left\{\begin{array}{ll}
      W(\mathbf{q}- z \mathbf{p}/
    \sqrt{n^2-|\mathbf{p}|^2},
    \mathbf{p},0) & ~~|\mathbf{p}|<n \\
      0 & ~~\mathrm{otherwise} \
    \end{array}\right. .
\end{equation}

\begin{figure}[p]
\caption{Propagation of the WDF in wide angle condition
($k=10^6~\mathrm{m}^{-1}$, horizontal scale in $\mu \mathrm{m}$).
a) Original distribution, b) after propagation over
$3~\mu\mathrm{m}$.}
    \vspace{20mm}
    \mbox{a) \psfig{file=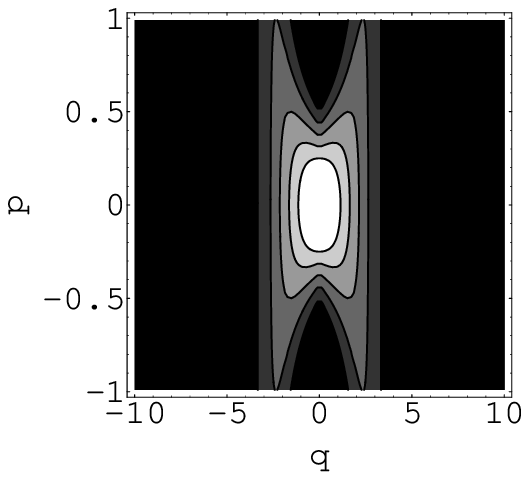, scale=1.5}}\\
    \mbox{b) \psfig{file=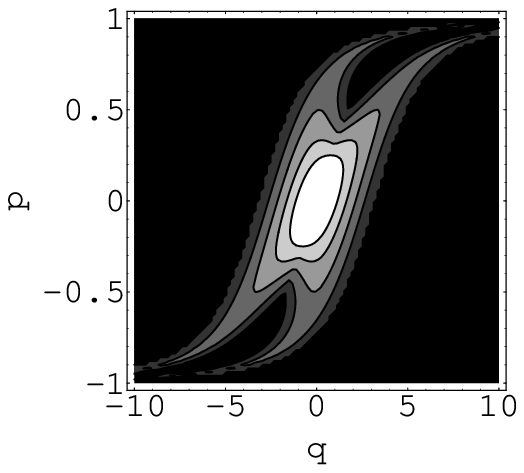, scale=1.5}}
\label{fig:wideshear}
\end{figure}

Eq.\ (\ref{eq:wideshear}) shows that only spatial frequencies
corresponding to momenta such that $|\mathbf{p}|<n$ can be
propagated \cite{Goodman68}. In fact $|\mathbf{p}|/n=\sin
\alpha$, with $\alpha$ the angle the ray makes with the $z$ axis.
It is then obvious that the higher frequencies would correspond to
values of $|\sin \alpha|>1$; these frequencies don't propagate and
originate evanescent waves instead, Fig.\ \ref{fig:wideshear}.
The net effect on near-field diffraction is that the
high-frequency detail near the aperture is quickly reduced.

The field intensity can now be evaluated by the expression
\begin{eqnarray}
 \label{eq:wideangle}
 \frac{k^2|\varphi(q,z)|^2}{4 \pi^2} &=& \int_{p_1}^{p_0}{\frac{1}{kp}
  \sin \left\{k p \left[l - 2 \left(q-z p / \sqrt{n^2-p^2}\right)\right]\right\}
   d p }\nonumber\\
 && +  \int_{p_0}^{p_2}{\frac{1}{kp}
  \sin \left\{k p \left[l + 2 \left(q-z p / \sqrt{n^2-p^2}\right)\right]\right\}
   d p } ,
\end{eqnarray}
with
\begin{eqnarray}
    p_0 &=& n q/\sqrt{q^2+z^2}, \nonumber \\
    p_1 &=& n(2q-l)/\sqrt{(2q-l)^2+4z^2} , \nonumber \\
    p_2 &=& n(2q+l)/\sqrt{(2q+l)^2+4z^2} .
\end{eqnarray}

\begin{figure}[p]
\caption{Near-field diffraction pattern when the aperture width
is exactly one wavelength; ($k=10^6~\mathrm{m}^{-1}$, both scales
in $\mathrm{\mu m}$).}
    \vspace{20mm}
    \centerline{\psfig{file=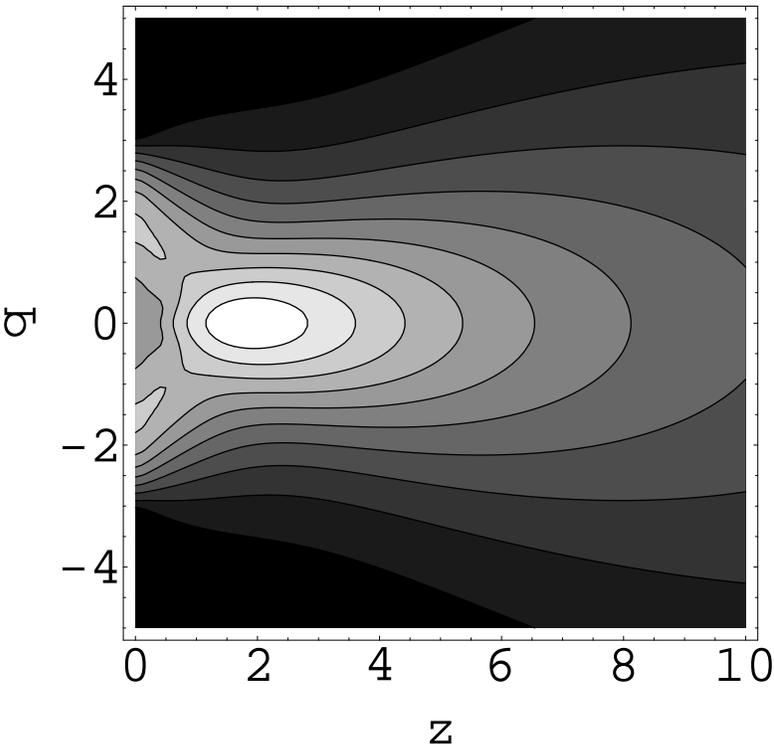, scale=2}}
\label{fig:wideangle}
\end{figure}

Fig.\ \ref{fig:wideangle} shows the near-field diffraction
pattern when the aperture is exactly one wavelength wide. The
situation is such that all the high order maxima of the WDF
appear at values of $|p|>1$ and are evanescent, resulting in a
field pattern with one small minimum immediately after the
aperture, after which the beam takes a quasi-gaussian shape,
without further minima. A sub-wavelength resolution would be
possible about half a wavelength in front of the aperture, where
the field distribution shows a very sharp peak.

We return now to the subject of FRFT to analyze the connection
between this and Fresnel diffraction. Being associated with a
rotation of the WDF, the FRFT can only represent diffraction if
associated with a quadratic phase factor that effectively
transforms the rotation in a shearing operation \cite{Alieva94}.
The implementation of the FRFT needs a combination of paraxial
free-space propagation, a thin lens and a magnification telescope
\cite{Agarwal94, Andres97}.

J. B. Almeida wishes to acknowledge the fruitful discussions with
P. Andr\'es, W. Furlan and G. Saavedra at the University of
Valencia.
 \pagebreak
  \bibliography{aberrations}   
  \bibliographystyle{OSA}   

\end{document}